\documentclass{emulateapj}
\usepackage{graphicx}

\shorttitle{Long Period Variables in the LMC}
\shortauthors{Fraser et al.}

\begin{document}

\slugcomment{AJ for publication in February 2005}

\title{Long Period Variables in the LMC: Results from MACHO and 2MASS}
\author{Oliver J. Fraser, Suzanne L. Hawley}
\affil{Department of Astronomy, University of Washington \\ Box 351580, Seattle WA, 98195-1580}
\email{fraser@astro.washington.edu, slh@astro.washington.edu}

\author{Kem H. Cook}
\affil{IGPP, Lawrence Livermore National Laboratory \\ MS L-413 \\ P.O. Box 808 \\ Livermore, CA 94550}
\email{kcook@llnl.gov}

\author{Stefan C. Keller}
\affil{Research School of Astronomy and Astrophysics \\ Australian National University \\ Mount Stromlo Obs. \\ Cotter Rd., Weston, ACT 2611, Australia}
\email{stefan@mso.anu.edu.au}
\begin{abstract}

We use the eight year light-curve database from the MACHO (MAssive Compact Halo Objects) project together with infrared colors and magnitudes from 2MASS (the Two Micron All Sky Survey) to identify a sample of 22,000 long period variables in the Large Magellanic Cloud (referred to hereafter as LMC LPVs). A period-luminosity diagram of these stars reveals six well-defined sequences, in substantial agreement with previous analyses of samples from OGLE (Optical Gravitational Lensing Experiment). In our analysis we identify analogues to galactic LPVs in the LMC LPV sample. We find that carbon-dominated AGB stars populate only two of the sequences, one of which includes the Mira variables. The high-luminosity end of the same two sequences are also the location of the only stars with $J-K_s > 2$, indicating that they are enshrouded in dust. The unknown mechanism that drives the variability of stars in the longest period produces different morphology in the period-luminosity diagram as compared to the shortest period sequences, which are thought to be caused by pulsation. In particular, the longest period sequence extends to lower luminosity RGB stars and the luminosity function does not peak among the AGB stars. We point out several features which will constrain new models of the period-luminosity sequences.

\end{abstract}


\section{Introduction}

Long period variables (LPVs) are giant stars that pulsate with periods ranging from weeks to years. The prototypical case is Mira (Omicron) Ceti, from which all large-amplitude ($>2.5$ magnitudes in V) regular LPVs take their name. LPVs are typically asymptotic giant branch (AGB) stars that are in the late stages of stellar evolution, where the pulsation mechanism(s) -- and the relationships between pulsation, mass loss, and the eventual ejection of the stellar envelope as the star becomes a planetary nebula -- are all poorly understood. During this relatively short AGB phase, stars manage to lose a significant fraction of their total mass, returning processed material to the interstellar medium. A qualitative picture of the mass-loss process invokes instabilities in the helium and hydrogen burning shells surrounding the degenerate core of the star. This provides the energy for a variety of stellar pulsation modes, causing shocks in the envelope which lift material high in the atmosphere.  The temperature is then low enough to allow the formation of dust, which is subsequently ejected from the atmosphere by radiation pressure from the luminous core.   Pulsations -- which affect both the radius and temperature of the star -- together with the shocks and dust formation, combine to produce significant photometric variability.

The study of LPVs has undergone a recent revival with the advent of surveys producing large catalogs of variable stars. Gravitational micro-lensing surveys, such as OGLE -- \cite{1994ApJ...435L.113P} and OGLE II -- \cite*{1997AcA....47..319U}; EROS -- \cite{1995A&A...301....1A}; MACHO -- \cite{macho}; and MOA -- \cite{2001MNRAS.327..868B} require that millions of stars be imaged nightly. The variable star catalogs that arise as a by-product contain thousands of stars with light curves covering several thousand days.

As noted by \cite{cook96} the long period variables in the MACHO database form five parallel sequences in period-luminosity space. \cite{wood99} classified these and suggested possible underlying mechanisms for the five. Our use of infrared magnitudes in the period-luminosity relations shown in Figure \ref{perlum} reveals six sequences. Wood's second sequence (``B'' in his notation) is split into our Sequences 2 and 3. In terms of our observed sequences, \cite{wood99} found that Sequence 1 corresponds to Mira type variables, which are theorized to be pulsating in the fundamental mode. Higher-order pulsations are invoked to explain the shorter period sequences (2, 3, and 4). The sequence that lies at longer-periods and lower luminosities than the Miras, E, showed light curves that suggested that all of these variable stars were eclipsing binaries.  Stars in the longest period sequence, Sequence D, proved more mysterious since the long secondary periods of these stars could not be explained by radial pulsations -- they are longer than the periods of Miras, which themselves are theorized to be pulsating in the fundamental mode. Additionally, they all exhibited multiple periods with the short period associated with Wood's Sequence ``B''. This prompted Wood et al. to propose that Sequence D was caused by the eclipse of the AGB star by a cloud of material around an unseen companion.

More recent studies using OGLE-II data [\cite{2002MNRAS.337L..31I}; \cite{kiss+bedding}; \cite{yoshi}; \cite{2004MNRAS.347L..83K}; \cite{Soszynski:2004wn}; \cite{Groenewegen:2004um}] have extended the work on the period-luminosity sequences. These authors showed that the use of infrared magnitudes (e.g. from 2MASS) for luminosity split the original second sequence into two sequences. Further, their data show that Wood's first two sequences (our first three sequences) exhibit a split in luminosity due to the contribution of red giant branch (RGB) stars. Most previous work on LMC LPVs overplot each period of these predominately multiperiodic stars. However, \cite{2004MNRAS.347L..83K} find there is no noticeable change in the period-luminosity sequences when plotting only the strongest period, or even when plotting the second strongest. For this reason results can be compared between those works and ones which find only a single period for each star: e.g. \cite{2002MNRAS.337L..31I}, \cite{yoshi}, and this paper.

In this paper we use the full eight-year MACHO database to re-examine the LPV period-luminosity relations.  MACHO observations have several advantages over previous studies: the time baseline of the light curves is twice as long (eight years compared to four years), and the colors of MACHO light curves yield superior sampling of photometric variability of RGB stars. As with previous studies, we use 2MASS infrared data to supplement our optical lightcurves.

\section{Data}
\subsection{The MACHO LMC Variable Star Database}
\label{macho_data}

The MACHO project \citep{macho} comprises eight years of observations of the Large and Small Magellanic Clouds and the Milky Way Bulge. In this work we select LMC  stars from the MACHO variable star catalog \citep{2003yCat.2247....0A}. Stars from the full database of several million stars were selected for this catalog if the central 80 percent of points in the light curve failed to fit a constant magnitude in a $\chi$-squared test.  This criterion resulted in 207,632 variable candidates in the LMC.\footnote{The MACHO variable star catalog is online at: \url{http://wwwmacho.mcmaster.ca/}}

MACHO employed a nonparametric phasing technique \citep{riemann} to find periods for each candidate variable star. Data were taken simultaneously in a red and a blue filter, and were analyzed independently to find a period and amplitude for each variable candidate. There is no substantial difference between the periods found, but a plot of the red versus the blue amplitudes shows a bias toward larger amplitudes in the blue. We adopt the blue periods and amplitudes so that the wider spread in amplitude will allow us to better discriminate types of LMC LPVs. The period-finding algorithm tends to alias noisy data and stars with chaotic variability. Thus, stars with period aliases at the total survey length, one year, and multiples of one day, in particular up to the fourth multiple and down to one-ninth ($\frac{1}{9}$) of a day, were all removed from the catalog. Note that these alias cuts result in blank vertical stripes in the period-luminosity diagram (Figure \ref{perlum}). Only 52 percent of the LPVs have well-determined periods, a total of 21,441 stars. We are currently analyzing the $\sim20,000$ LPVs with poorly-determined periods with the goal of understanding the multi-periodic nature of these stars.

\label{amplitudes}
The amplitudes in the MACHO database are determined from the difference between the median of five points nearest the maximum and five points nearest the minimum in the light curve. A model light curve is used to identify the times of maximum and minimum light, so stars that are not well fit by the model will tend to average to zero amplitude. Thus, noisy or chaotic data will not be assigned meaningful amplitudes.

\subsection{2MASS Photometry}
\label{2mass}

The Two Micron All Sky Survey (2MASS) measured $J$, $H$, and $K_s$ magnitudes for approximately half a billion objects over the entire sky \citep{2mass}. The sky coverage, resolution, and sensitivity of 2MASS make it a very useful source of infrared magnitudes. We have matched the MACHO and 2MASS data by position for LMC LPVs with well-determined periods. Since the 2MASS magnitudes were taken at random phases we expect scatter in the $K$ magnitudes from the intrinsic variability of these LPVs. Although most of the stars in our sample are small amplitude pulsators and contribute a negligible amount of scatter in the $K_s$ band, some of the stars that make up Sequence 4 are Mira variables. Miras are the largest amplitude LPVs and have a typical $K$ amplitude of $\sim 0.6$ \citep{2000PASA...17...18W}. Thus the 2MASS magnitudes of the largest amplitude stars may deviate from the mean by as much as 0.3 mags. We accept this as a source of scatter, and, assuming all LMC stars are at the same distance, we use the 2MASS $K_s$ magnitude as our primary luminosity indicator.

\begin{figure*}[t]
\begin{center}
\emph{The figures have been commented out of the \LaTeX source as I couldn't make postscript files that were both small and sharp. The png files I've uploaded instead look great and are nice and small. You can see them as I intended if you uncomment the \texttt{\\includegraphics} commands in the source and run pdflatex a few times.}
\caption{LMC period-luminosity diagram showing the stars with well-determined periods.  The luminosity split at the top of the RGB ($K_s = 12.3\pm 0.1$) is observed in sequences 2, 3, and 4. }
\label{perlum}
\end{center}
\end{figure*}

Infrared magnitudes can also be used to distinguish the state of late AGB stars. After evolving through the RGB and horizontal branch, stars begin to ascend the AGB with a high O/C ratio in their atmospheres. These oxygen-rich AGB stars include both the so-called early AGB stars and AGB stars that have had their first thermal pulses. Low mass stars will remain on the AGB long enough that thermal pulses initiate the ``third dredge-up'', the deepening of their convection zones down into material enriched in $^{12}$C. The spectral type then changes to a carbon star. The objective-prism survey of the LMC by \cite{2001A&A...369..932K} showed that carbon stars appear on the infrared color-color diagram in a red tail running from the stellar locus. Infrared color cuts of $J-H>0.89$ and $H-K_s>0.32$ isolated 83\% of the carbon stars in their sample, with only a nine percent contamination rate. A single color cut in $J-K_s$ can also select these stars since it is perpendicular to the tail of carbon stars; the result is a sample that is less complete but also has less contamination. In addition, the 2MASS color-magnitude diagram of the LMC \citep{2000ApJ...542..804N} shows a clear division at $J-K_s = 1.4$ due to carbon stars. In this paper we will draw particular attention to LPVs near this color boundary with the caveat that we are missing some twenty percent of the carbon stars -- those that are warm and hence bluer than $J-K_s = 1.4$.

\cite{2000ApJ...542..804N} also find a population of very red stars ($J-K_s > 2$) that lie along the reddening vector from the carbon stars. They propose that these stars are surrounded by a dusty circumstellar envelope. Both AGB stars and at least a few protostars are known to lie in this region. Although such dusty LPV are likely to be very late AGB stars, pin-pointing the exact evolutionary state of these latter stars will require a better understanding of the connection between mass-loss, pulsation, and the late stages of AGB star evolution.

\section{Discussion}

Period-luminosity relations for variable stars in the LMC are shown in Figure \ref{perlum}. The two Cepheid sequences are clearly visible between $0 <$ log P $< 1$ and $13 < K_s < 16$. When plotted in $K_s$, as shown here, it is evident that Wood's original sequences split in both period and luminosity. It also appears that the low luminosity ends of the three shortest period sequences are offset to longer periods. These effects have been previously seen in OGLE II data; \cite{kiss+bedding}, \cite{yoshi}. The lower luminosity stars in Sequences 2, 3, and 4 have been identified as RGB stars \citep{2004MNRAS.347L..83K} using an analysis of the second derivative of the luminosity function. The number of stars located above and below the tip of the RGB in each pulsation sequence is given in Table \ref{counts}. For the tip of the RGB we use $K_s = 12.3 \pm 0.1$ from \cite{2000ApJ...542..804N} who analyzed the luminosity function of all 2MASS stars in the direction of the LMC. Note that some previous authors have used $K_s = 12.1$, found by analyzing the luminosity function of only the variable stars.

\begin{deluxetable}{ccccc}
\tablecaption{Properties of the Period-Luminosity Sequences\label{counts}}
\tablehead{
\colhead{Sequence}
& \multicolumn{2}{c}{Star Count}
& \multicolumn{2}{c}{Amplitude}\\
& \colhead{above TRGB\tablenotemark{a}}
& \colhead{below TRGB\tablenotemark{a}}
& \colhead{Mean\tablenotemark{b}}
& \colhead{Max}
}
\startdata
4&1337&673&0.11&1.75\\
3&2780&540&0.17&1.90\\
2&3967&310&0.34&3.79\\
1&5009&628&0.77&8.04\\
E&636&1384&0.28&4.47\\
D&2163&1802&0.38&5.58\\
\enddata
\tablenotetext{a}{Tip of the Red Giant Branch; $K_s = 12.3\pm 0.1$ in the LMC \citep{2000ApJ...542..804N}. Above the tip of the RGB the LPVs are all AGB stars.}
\tablenotetext{b}{We are unable to estimate a minimum amplitude of pulsation as MACHO's amplitude finding algorithm reports \emph{zero} amplitude for noisy or chaotic data (see \S \ref{amplitudes}).}
\end{deluxetable}

\subsection{Connections Between LMC and Galactic LPVs}

Galactic LPVs are identified primarily in The General Catalog of Variable Stars \citep{GCVS} which has traditionally classified LPVs by their $V$ band behavior. The classification scheme, tabulated in Table \ref{lpv-types}, separates LPVs into Mira types and classes of semi-regular (SR) variables. Mira variables are defined as those pulsating with a period greater than 80 days and an amplitude of at least 2.5 in $V$. Whereas Mira variables show regular and strong periodicity, one would imagine that ``semi-regular'' variables would have weak or poorly defined periods. Indeed, SRb stars have poorly expressed periodicity, \emph{or} multiple periods, \emph{or} only occasional periodicity, \emph{or} chaotic pulsation. However, SRa stars are not necessarily semi-regular at all; they are similar to Miras but with smaller amplitudes \citep*{2002A&A...393..573L}. 

Due to the empirical nature of the GCVS classification system, connections drawn to the observed LMC LPV period-luminosity sequences are often not simple or direct. However, connections can be made; \cite{2002A&A...393..573L} find that Miras in the Galaxy and in the Magellanic Clouds have the same period-luminosity relationship, implying that the pulsation mechanism is not strongly dependent on metallicity. Also, parallel period-luminosity sequences exist in Galactic samples (\cite{1998ApJ...506L..47B}; \cite{2003A&A...403..993K}), as in the LMC.

\begin{figure*}
\begin{center}
\caption{
Period-Luminosity diagrams highlighting stars in seven amplitude bins (with divisions at $B_{MACHO} =$ 0.07, 0.17, 0.3, 0.7, 1, and 2.5).  The background distribution (in gray) is for all of the LMC LPVs from Figure \ref{perlum} and is shown as a reference. Note that the highest amplitude stars are the Mira variables, located on the long period side of Sequence 1.}
\label{amp}
\end{center}
\end{figure*}

\begin{figure*}
\begin{center}
\caption{
Period-Luminosity diagrams highlighting stars in five $J-K_s$ color bins (with divisions at $J-K_s$ = 1.1, 1.2, 1.4, and 2). The background distribution (in gray) is for all of the LMC LPVs from Figure \ref{perlum} and is shown as a reference. Note that stars with $J-K_s > 1.4$ are carbon stars, and stars with $J-K_s > 2$ are obscured by dust (see \S \ref{2mass}). The bluest stars ($J-K_s < 1.1$) show the RGB populations of the period-luminosity sequences.}
\label{jk}
\end{center}
\end{figure*}

\begin{deluxetable*}{lllcl}
\tablecaption{GCVS Long Period Variable Classification\label{lpv-types}}
\tablehead{
\colhead{Type}
& \colhead{Spectral Type}
& \colhead{Period (days)}
& \colhead{Amplitude (V)}
& \colhead{P-L sequence(s)}
}
\startdata
Mira&late\tablenotemark{a} giants&80 -- 1000&$\ge2.5$&4\\
SRa&late\tablenotemark{a} giants &35 -- 1200&$<2.5$&1,2,3,4\\
SRb&late\tablenotemark{a} giants&20 -- 2300&\nodata&1,2,3,6\\
SRc\tablenotemark{b}&late supergiants&30 -- ``several thousand''&$\sim1$&\nodata\\
SRd\tablenotemark{b}&F, G, \& K&30 -- 1100&0.1 -- 4&\nodata\\
\enddata
\tablenotetext{a}{a ``late'' giant specifically indicates a spectral type of M, C, S, Me, Ce, or Se.}
\tablenotetext{b}{These rare types are not represented in our collection of late giants.}
\end{deluxetable*}

\begin{figure*}
\begin{center}
\caption{Color-magnitude diagrams and luminosity functions (along the left vertical axes) for each period-luminosity sequence arranged from shortest to longest period (see Figure \ref{perlum}). The background distribution in the color-magnitude diagram is for the entire MACHO variable star catalog, whereas the background luminosity function is for \emph{all} LMC stars measured by 2MASS \citep{2000ApJ...542..804N}. The tip of the RGB is marked at $K_s = 12.3$ in each panel.}
\label{cmd}
\end{center}
\end{figure*}

 Figure \ref{amp} shows LMC LPVs of different amplitudes located in period-luminosity space, and Table \ref{counts} gives the mean and maximum pulsation amplitudes for each sequence. There is only a small difference between our $B_{MACHO}$ amplitudes and the $V$ band amplitudes used to classify Galactic variables. \cite{1999PASP..111.1421B} found transformations from MACHO magnitudes to Cousins $V$ and $R$. For $B_{MACHO}-R_{MACHO} > 1.0$ 

\begin{equation}
V - B_{MACHO} = -0.07 - 0.10(B_{MACHO}-R_{MACHO}).
\end{equation}

The change in the amplitude is thus due to color changes during the star's variability cycle. In particular, the change in the amplitude between $V$ and $B_{MACHO}$ is restricted to 10\% of the color change in the MACHO photometric system.\footnote{Since we know the typical color of AGB stars, $B_{MACHO}-R_{MACHO} \sim 1.5$ \citep{1999PASP..111.1539A}, we note that $V$ magnitudes will be $\sim0.22$ brighter than the $B_{MACHO}$ magnitudes.}

With reference to the LMC LPVs in Figure \ref{amp}, Sequence 1 is composed on the long period side by Miras. \cite{2001A&A...377..945C} found that higher amplitude SRa stars lie on Sequence 1 with the Miras; low amplitude SRas, on the other hand, lie on the shorter period sequences (Sequences 2, 3, and 4). \cite{2001A&A...377..945C} investigate stars from Sequence D and find that they are all multiply-periodic SRb stars. \cite{wood99} found that the shorter period of these stars usually fell on his sequence ``B'', which corresponds to our Sequences 2 and 3. Wood's work did not identify any stars from Sequence D with a primary period on Sequence 4. Unlike sequence 1, the highest amplitude stars fall on the shorter period side of the sequence. 

The RGB stars at the low luminosity end of Sequences 3 and 4 are low amplitude pulsators, and a similar set of stars is visible at the base of the Sequence D. The latter sequence's stars are particularly interesting as their variability mechanism remains unknown. Sequence D is also notable for a significant number of high amplitude variables ($1 < B_{amp} < 2.5$).

\subsection{Color and Luminosity Functions}
\label{cmd_section}

Figure \ref{jk} shows LMC LPVs of different $J-K_s$ color located on the period-luminosity diagram; as discussed in \S \ref{2mass}, stars with $J-K_s > 1.4$ are likely to be carbon stars, while stars with $J-K_s > 2$ are likely obscured by dust. Interestingly, the most heavily obscured stars lie only at the high luminosity end of Sequence 1 and are Miras. The majority of carbon stars lie at the high luminosity end of Sequences 1 and 2. Sequence D shows that the lower luminosity, small amplitude pulsators visible in Figure \ref{amp} also have bluer colors than the rest of the sequence, again indicating a separate RGB population.

The color-magnitude diagrams and luminosity functions for each sequence are shown in Figure \ref{cmd}; the luminosity functions are scaled such that the relative population of stars in each sequence can be compared. The background distribution in the color-magnitude diagram is for the entire MACHO variable star catalog, whereas the background luminosity function is for \emph{all} LMC stars measured by 2MASS \citep{2000ApJ...542..804N}.

The ``bumps'' in the luminosity function for each sequence correspond to particular populations at different stages of stellar evolution. The top of the RGB is the largest such bump, the tip of which was found by \cite{2000ApJ...542..804N} to be $K_s = 12.3 \pm 0.1$. Sequences 3 and 4 show a substantial contribution from the top of the RGB, whereas Sequences 1 and 2 are composed primarily of stars on the AGB (which overlaps with the RGB at this magnitude). Sequences 1, 2, 3, and 4 all show a peak in the luminosity function at $K_s \sim 11.3$; this peak in variable AGB stars is not seen in the overall LMC luminosity function \citep{2000ApJ...542..804N}.

The types of stars in each sequence are obtained by comparison with the color-magnitude diagram of the LMC from \cite{2000ApJ...542..804N}. Sequences 3 and 4 have similar distributions including a population at the tip of the RGB as well as oxygen-rich AGB stars. The stars at $J-K_s \approx 1.25$ with $K_s < 10.25$ are young, massive AGB stars; they are most commonly found in Sequence 3. Sequences 1 and 2 also have similar populations, with both oxygen-rich AGB stars and a large contribution from carbon stars. Sequence 1 extends further to the red in $J-K_s$, probably due to dust obscuration around those stars. Sequence E is dominated by RGB stars, as can be seen from both its luminosity function and position in the color-magnitude diagram. This is expected if it is composed of eclipsing binaries. Both Sequence D and E lack high luminosity young, massive AGB stars and the AGB dominated population that produces the significant peak in the luminosity function at $K_s \sim 11.3$ in Sequences 1, 2, 3, and 4. Interestingly, Sequence D ends abruptly at $K_s \sim 13.7$ (Sequence E is artificially cut off at $K_s = 14.5$). There is no feature in the ``deep'' LMC color-magnitude diagram \citep{2000ApJ...542..804N} associated with this position, or with the bottom of the first four sequences. The difference in the low-luminosity cut-off of Sequence D compared with the first four sequences, as well as the lack of a bump in the luminosity function at $K_s \sim 11.3$ in Sequence D, suggests that the variability is caused by a different mechanism than the radial pulsation modes proposed to explain Sequences 1, 2, 3, and 4.

\begin{figure}
\begin{center}
\caption{Period-luminosity diagram overlaid with the observed LMC Mira relation of \cite{1989MNRAS.241..375F} (dashed line) and the 3rd, 2nd, and 1st overtone (respectively, from left) models of \cite{1996MNRAS.282..958W} (solid lines). The fundamental mode of the models was forced to fit the Mira relation of \cite{1989MNRAS.241..375F}.}
\label{theory}
\end{center}
\end{figure}

\subsection{Comparisons to Theory}

Explanations of the structure in the period-luminosity diagram often invoke radial pulsations, with parallel sequences indicating populations of overtone pulsators. We have plotted the linear, radial, non-adiabatic pulsation models of \cite{1996MNRAS.282..958W} with our LMC LPV data in Figure \ref{theory}. These models were forced to fit the observed LMC Mira period-luminosity relation of \cite{1989MNRAS.241..375F}. Note that the observed Mira relation lies precisely where we see the largest amplitude pulsators in Figure \ref{amp}. Sequences 1, 2, 3, and 4 have all been previously attributed to overtone pulsations, but Sequence 4 seems to lie apart from the rest. Sequences 3 and 4 share every important feature discussed in this paper -- morphology in the color-magnitude diagram, luminosity function, infrared colors, and, to a lesser degree, amplitudes (see Table \ref{counts}). New models are required in order to decide whether Sequence 4 represents the next highest overtone pulsation, or if there are damped pulsation modes between Sequences 3 and 4. Sequence 4 may not be shortest period-luminosity sequence as \cite{Soszynski:2004wn} find evidence of a sparcely populated sequence at still shorter period.

Sequence D was originally theorized by \cite{wood99} to be caused by eclipses from a dust-enshrouded unseen companion. This hypothesis required that $\sim 25 \%$ of AGB stars exist in semi-detached binaries. \cite*{2004ApJ...604..800W} have recently completed an analysis of four years of echelle data for three Galactic stars with long secondary periods (examples of which are tabulated in \cite{1963AJ.....68..253H}). Additional photometric data for 111 stars from MACHO allow them to investigate, and rule out, explanations involving solely radial pulsations, non-radial pulsations, orbiting companions, non-spherically symmetric stars, dust obscuration, and chromospheric activity. They propose that the most likely explanation is a low-degree $g^+$ mode in an extraordinarily thick radiative layer, which would allow large amplitudes at the stellar surface, combined with large-scale star spot activity. \cite{Soszynski:2004wn} find that Sequence D is made up of two populations: stars that do not have one of their periods on sequence 4 and stars that do. This split seperates Miras and SR variables from very small amplitude red variables respectively. Thus Sequence D is composed of stars which cannot be uniquely assigned to any specific class (AGB, RGB, Mira, SRa, SRb, very small amplitude pulsators, etc.).

\section{Conclusions and Future Work}

The combination of observations over a long time baseline, accurate optical photometry extending below the tip of the RGB, and 2MASS infrared photometry has allowed us to place multiple constraints on the stars inhabiting each of the sequences observed in the period-luminosity diagram of the LMC. 

\begin{itemize}
\item Most of the LPVs in this analysis, which comprises the 52\% of LMC LPVs with well-determined periods, are SRa stars or Miras. The amplitude criterion ($\delta V > 2.5$ for Miras) used to differentiate Galactic SRa and Mira stars seems arbitrary, as SRa's and Miras overlap in our diagrams. Stars from Sequence D are SRb stars as previous work has found that they are never monoperiodic.

\item Our period-luminosity Sequences 3 and 4 both show low luminosity extensions comprised of RGB stars, and have AGB stars that are oxygen rich ($J-K_s < 1.4$). As discussed in \S \ref{2mass} this implies that these AGB stars haven't yet undergone the third dredge-up. Compared with the significant population of redder (carbon) stars in Sequences 1 and 2 this implies that older stars are segregated to the latter two sequences. The populations of young, massive AGB stars in Sequences 3 and 4 that are identified in Figure \ref{cmd} (see \ref{cmd_section}) support this hypothesis.

\item The shortest period sequence (Sequence 4) is significantly seperated in period from the longer period sequences but it is otherwise undistinguished.

\item Sequences D and E tend to follow the LMC luminosity function more closely than the others, indicating that they are more nearly drawn from the general population of giant stars in the LMC. In particular, neither sequence shows a ``bump'' in its luminosity function due to AGB stars, unlike Sequences 1, 2, 3, and 4. These characteristics suggest that a different mechanism produces the variability of these sequences. Sequence E is likely populated by eclipsing binary systems.

\item Although the physical mechanism that causes Sequence D is still elusive, there are numerous constraints on possible models. \cite{Soszynski:2004wn} show that this sequence is composed of bright high amplitude Mira and SR variable stars \emph{and} a dimmer population of small amplitude red variables. This latter population is bluer, extends to lower luminosity, and is broader in period than the RGB star components of Sequences 2, 3, and 4.t The AGB population in Sequence D exhibits higher amplitude pulsation on its shorter period side, opposite what is found for Sequence 1.

\end{itemize}

We are currently analyzing the $\sim20,000$ multiply periodic LMC LPVs that are not included in the present sample, with the goal of understanding the relationships between the period-luminosity sequences and the positions in period-luminosity space of stars at various stages of their evolution.

\acknowledgments

This paper utilizes public domain data obtained by the MACHO Project, jointly funded by the US Department of Energy through the University of California, Lawrence Livermore National Laboratory under contract No. W-7405-Eng-48, by the National Science Foundation through the Center for Particle Astrophysics of the University of California under cooperative agreement AST-8809616, and by the Mount Stromlo and Siding Spring Observatory, part of the Australian National University.

This work was performed under the auspices of the U.S. Department of Energy, National Nuclear Security Administration by the University of California, Lawrence Livermore National Laboratory under contract No. W-7405-Eng-48.

This publication makes use of data products from the Two Micron All Sky Survey, which is a joint project of the University of Massachusetts and the Infrared Processing and Analysis Center/California Institute of Technology, funded by the National Aeronautics and Space Administration and the National Science Foundation.

Oliver Fraser and Suzanne Hawley acknowledge support by NSF grant AST 02-05875.

\bibliography{../refdb}

\end{document}